\documentclass[preprint,nofootinbib]{revtex4}
\usepackage{amsmath}
\usepackage{amsfonts}
\usepackage{amssymb}
\usepackage{graphicx}

\newcommand{\Te}{\mathbb{T}}

\newcommand{\Opd}{{\cal O}(p^2)}

\begin{document}

\title{Is there an isovector companion of the $X(2175)$?}
\author{C. A. Vaquera-Araujo\footnote{Email:vaquera@fisica.ugto.mx} and M. Napsuciale\footnote{Email:mauro@fisica.ugto.mx}}
\affiliation{ Departamento de F\'{\i}sica, Divisi\'{o}n de Ciencias
e Ingenier\'{\i}as, Universidad de Guanajuato, Campus Le\'{o}n;
Lomas del Bosque 103, Fraccionamiento Lomas del Campestre, 37150,
Le\'{o}n, Guanajuato, Mexico.}

\begin{abstract}
In this letter we study the reaction $e^{+}e^{-}\rightarrow\phi \pi^0 \eta$
with the $\pi^{0}\eta$ system in an isoscalar s-wave configuration. We use a
formalism recently developed for the study of $e^{+}e^{-}\rightarrow\phi \pi\pi$
and  $e^{+}e^{-}\rightarrow\phi K^{+}K^{-}$. The obtained cross section is within
the reach of present $e^{+}e^{-}$ machines. Measuring this channel would test the
absence of an isovector companion of the $X(2175)$ as predicted by the three body
approach to this resonance.
\end{abstract}


\maketitle


\section{Introduction}
Recently,  a detailed study of the reactions
$e^+e^- \to \phi \pi\pi$ and $e^+e^- \to \phi K^{+}K^{-}$
was performed  for $\sqrt{s}\geq 2$ GeV \cite{Napsuciale:2007wp, GomezAvila:2007ru}.
These processes involve the $\gamma\phi \pi\pi$ and $\gamma\phi K^{+}K^{-}$ vertex
functions respectively which are calculated at low photon virtuality and low di-meson
invariant mass using Resonance Chiral Perturbation Theory ($R\chi PT$). The dynamics
is shown to be dominated by electromagnetic form factors and meson-meson amplitudes.
This calculation, valid for low intermediate photon virtuality is improved replacing
the lowest order terms in the form factors by the
full form factors at the required $\sqrt{s}$ as extracted from experiment or by the
unitarized form factors. Likewise,
the lowest order terms in the meson-meson amplitudes are replaced by the unitarized
meson-meson amplitudes containing the scalar poles. The related process
$e^{+}e^{-}\rightarrow\phi f_{0}$ has been also studied in the context of
Nambu-Jona-Lasinio models \cite{Kuraev}.

In the $e^+e^- \to \phi \pi\pi$ case, the calculation reproduces the
background in the total cross section but not the peak around $2175$ MeV,
discovered by BaBar \cite{Aubert} and confirmed by
Bes in $J/\Psi \to \eta \phi f_0(980)$ \cite{:2007yt} and Belle also
in $e^+e^- \to \phi \pi^{+}\pi^{-}$ \cite{:2008ska}.

The BaBar values for the mass and width of this new resonance, known
as $X(2175)$ or $Y(2175)$, are $M_X = 2175 \pm 10\, \text{MeV}$ and
$ \Gamma_X = 58 \pm 16 \pm 20\, \text{MeV}$  \cite{Aubert}, which
are consistent with Bes, $M_X = 2186 \pm 10 \pm 6\,
\text{MeV}$ and $ \Gamma_X = 65 \pm 23 \pm 17\,  \text{MeV}$
\cite{:2007yt}, and Belle,
$M_X = 2079\pm 13 ^{+79}_{-28}\, \text{MeV}$ and
$ \Gamma_X = 192 \pm 23 ^{+25}_{-61}\,  \text{MeV} $ \cite{:2008ska} results.
As pointed out in a recent review
\cite{Zhu:2007wz}, the narrow width is at odds with the large decay
width into two mesons predicted in  models for $\bar{s}s$ states
\cite{Barnes:2002mu}, diquark-anti-diquark states
\cite{Ding:2006vk}, tetraquark $s\bar{s} s \bar{s}$ states
\cite{Wang:2006ri} or  gluon hybrids $s \bar{s} g$
\cite{Ding:2007pc} (see also \cite{Close:2007ny}).

Another proposal for the nature of $X(2175)$, which turns out to be
consistent with the observed mass and width, is a three-body $\phi K
\bar{K}$ system \cite{MartinezTorres:2008gy}. The result of this
framework is a neat resonance peak around a total mass of $2150\,
\text{MeV}$ and an invariant mass for the $K\bar{K}$ system around
$970\, \text{MeV}$, quite close to the $f_0(980)$ mass. The state
appears in the isospin $I=0$ sector, and qualifies as a $\phi
f_0(980)$ resonance. Interestingly,  the theory also shows that
there is no resonance in the isovector channel. This is the main
motivation to study the reaction $e^+ e^- \to \phi \pi^{0}\eta $,
whose final state is in a pure isovector state.

\section{Calculation of $e^{+}e^{-}\rightarrow\phi\pi^0\eta$ }
In this work we apply the same formalism as in \cite{Napsuciale:2007wp, GomezAvila:2007ru}
to the production of $\phi \pi^{0}\eta $. We start form the $R\chi PT$ Lagrangian and
follow the conventions in \cite{EGPR}. The relevant
interactions in their notation are%
\begin{eqnarray}
\mathcal{L} & =&\mathcal{L}^{(2)}+\mathcal{L}^{(F)}+\mathcal{L}^{(G)}
\label{lag} \\
\mathcal{L}^{(2)} & =&\frac{1}{4}f^{2}tr\left( \left( D_{\mu}U\right)
^{\dagger}D^{\mu}U+\chi U^{\dagger}+\chi^{\dagger}U\right)  \label{L2} \\
\mathcal{L}^{(F)} & =&\frac{F_{V}}{2\sqrt{2}}tr(V_{\mu\nu}f_{+}^{\mu\nu })
\label{LF} \\
\mathcal{L}^{(G)} & =&\frac{iG_{V}}{\sqrt{2}}tr(V_{\mu\nu}u^{\mu}u^{\nu}),
\label{LG}
\end{eqnarray}
where
\begin{eqnarray}
u_{\mu} & =&iu^{\dagger}\ D_{\mu}U\ u^{\dagger},\qquad U=u^{2},\qquad u=e^{-%
\frac{i}{\sqrt{2}}\frac{\Phi}{f}},\qquad\Phi=\frac{1}{\sqrt{2}}%
\lambda_{i}\varphi_{i}  \label{umu} \\
f_{+}^{\mu\nu} & =&u\ F_{L}^{\mu\nu}u^{\dagger}+u^{\dagger}\ F_{R}^{\mu\nu }\
u,\qquad D_{\mu}U=\partial_{\mu}U-i\left[ v_{\mu},U\right] .
\end{eqnarray}
We introduce the photon field through $v_{\mu}=eQA_{\mu}$\ and $%
F_{L}^{\mu\nu }=F_{R}^{\mu\nu}=eQF^{\mu\nu}$ ($e>0$) where $F^{\mu\nu}\ $%
denotes the electromagnetic strength tensor.

\begin{figure}[ptb]
\begin{center}
\includegraphics[width=0.6 \textwidth ]{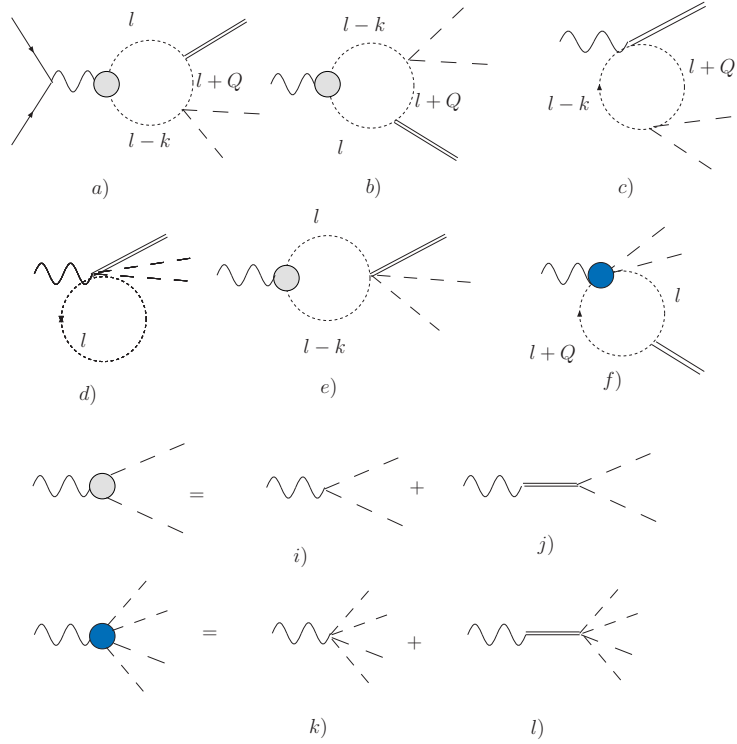}
\end{center}
\caption{Feynman diagrams for $e^{+}e^{-}\rightarrow\protect\phi\protect\pi^0%
\protect\eta$ in $R\protect\chi PT$. See \cite{Napsuciale:2007wp} for
the details and conventions} \label{FD}
\end{figure}

There are no tree-level contributions to $e^{+}e^{-}\rightarrow\phi\pi^0\eta$ in $R\chi PT$.
At one loop level, this process is induced by the diagrams shown in Fig. (\ref{FD}) where
the fermionic lines are shown only in the first diagram and particles in the loops are
neutral and charged kaons. For the sake of simplicity a shaded circle and a dark
circle account for the diagrams $i)$ plus $j)$ and $k)$ plus $l)$
respectively, which differentiate the direct photon coupling from the
coupling through an intermediate vector meson. We will address the
corresponding diagrams as $a)$, $b)$, when we have the direct photon
coupling and $a^{\prime})$, $b^{\prime})$, when the coupling goes through
the exchange of a vector meson.

\begin{figure}[ptb]
\begin{center}
\includegraphics[width=0.4 \textwidth ]{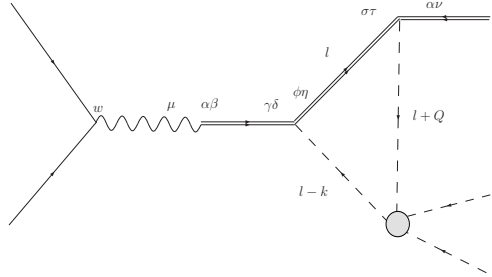}
\end{center}
\caption{Feynman diagram for $e^{+}e^{-}\to K^{*}\bar{K}\to
\protect\phi K \bar{K}\to \protect\phi \protect\pi^0\protect\eta$. }
\label{FDV}
\end{figure}

At the energy of the reaction,$\sqrt{s}\geq 1.7 $GeV, it is quite probable
to excite higher states. Here, we consider the excitation of virtual $K^{\ast}K$ states
and their contribution to $e^{+}e^{-}\rightarrow\phi\pi^0 \eta$ through the chain
$e^{+}e^{-}\rightarrow K^{\ast} \overline{K}\rightarrow\phi K \overline{K}$ with the
rescattering of the kaon pair to $\pi^0\eta$ as shown in Fig. (\ref{FDV}).

The $VV^{\prime}P$ interaction in this diagram is dictated by the anomalous
chiral Lagrangian \cite{Bijnens:1989jb} which we rewrite in terms of the tensor
field as
\begin{equation}
\mathcal{L}_{anom}=\frac{G}{\sqrt{2}}\epsilon_{\mu\nu\alpha\beta}tr(%
\partial^{\mu}V^{\nu}\partial^{\alpha}V^{\beta}\Phi)=\frac{G_{T}}{4\sqrt {2}}%
\epsilon_{\mu\nu\alpha\beta}tr(V^{\mu\nu}V^{\alpha\beta}\Phi).
\label{anom}
\end{equation}
with $G_{T}=M_{V}M_{V^{\prime}}G$. This Lagrangian was obtained in \cite{Bijnens:1989jb}.
In this formalism no direct $VP\gamma$ coupling emerges and this vertex is generated by the Lagrangian in Eq. (\ref{anom}) in combination with the $V\gamma$ interaction in Eq. (\ref{LF})
(see the discussion above Eq. (56) of \cite{Bijnens:1989jb}).

The calculation of the invariant amplitude is similar to the
previously studied $\phi\pi\pi$ final state and we refer the reader
to Ref. \cite{Napsuciale:2007wp} for the details. Both $R\chi PT$
contributions in Fig. (\ref{FD}) and the contributions of Fig.
(\ref{FDV}) turn out to be divergent. However, it is shown in
\cite{Napsuciale:2007wp} that divergences of $R\chi PT$
contributions come from diagram {\it c)} of Fig. (\ref{FD}) and
involve the same divergent scalar integral which appears in the
unitarization of the meson-meson amplitudes. On the other side, a
careful decomposition of the loop integrals of Fig. (\ref{FDV}) into
scalar integrals was performed in the appendix of Ref.
\cite{Napsuciale:2007wp}. Using dimensional regularization it is
shown there that the only divergent piece corresponds to the same
scalar integral appearing in meson-meson scattering. The
substraction constant required for this integral was discussed in
\cite{OO,OOPel} and we briefly review it below.

The total amplitude for $e^{+}(p^{+})e^{-}(p^{-})\to \phi (Q,\epsilon) \pi^0 (p) \eta (p^{\prime})$ arising from the diagrams in Figs. (\ref{FD},\ref{FDV}) is
\begin{equation}
-i\mathcal{M}=\frac{ie^{2}}{2\pi ^{2}m_{K}^{2}}
\frac{V_{KK\pi\eta}}{\sqrt{2}}\frac{L^{\mu}}{k^{2}}
\left[ I\ L_{\mu \nu }^{(1)}-J\ L_{\mu \nu }^{(2)}\right]
\epsilon^{\nu }  \label{finalamp}
\end{equation}%
where $k^{2}=(p^{+}+p^{-})^{2}$, $V_{KK\pi\eta}$ denotes the leading order on-shell
amplitude for $K\overline{K}-\pi^0\eta$ scattering, $L^{\mu}\equiv\overline{v}%
(p^{+})\gamma^{\mu }u(p^{-})$ and the Lorentz covariant structures are given by
\begin{equation}
L_{\mu \nu }^{(1)}\equiv Q\cdot kg_{\mu \nu }-Q_{\mu }k_{\nu },\qquad L_{\mu
\nu }^{(2)}=k^{2}g_{\mu \nu }-k_{\mu }k_{\nu }.  \label{L12}
\end{equation}
The $I,\, J$ functions entering Eq. (\ref{finalamp}) are given by
\begin{eqnarray}
I& =&\frac{\sqrt{2}M_{\phi }}{2f^{2}} G_{V}F_{K}^{(1)}(k^{2})I_{P}
-\frac{G}{4\sqrt{2}}F_{K^{\ast }K}^{(1)}(k^{2}) \times \nonumber \\
&&\left\{ Q\cdot k J_{V}
-m_{K}^{2}\left\{ I_{G}-I_{2}+4-\frac{1}{2}[a(\mu)+3]
+\frac{1}{2}\log \frac{m_{K}^{2}}{\mu ^{2}%
}\right\} \right\} \label{Idecay} \\
J& =&\frac{\sqrt{2}M_{\phi }G_{V}}{2f^{2}}F_{K}^{(1)}(k^{2})\left[ J_{P}+\frac{%
(4\pi m_{K})^{2}}{4k^{2}}G_{KK}(m^2_{\pi \eta})\right] \nonumber \\
&&-\frac{GM_{\phi}^{2}}{4\sqrt{2}}F_{K^{\ast }K}^{(1)}(k^{2})J_{V},  \label{Jdecay}
\end{eqnarray}%
with the divergent loop integral
\begin{eqnarray}
G_{KK}(m_{\pi\eta}^{2})&=&\mu^{2\varepsilon}\int\frac{d^{d}l}{(2\pi)^{d}}\frac {i%
}{\square_K\left( l+Q\right) \square_K\left( l-k\right) } \nonumber \\
&=&\frac{1}{\left(
4\pi\right) ^{2}}\left[ a(\mu)+\log\frac{m_{K}^{2}}{\mu^{2}}+I_{G}(m_{\pi
\eta}^{2})\right] , \label{GKreg}
\end{eqnarray}
where $\mu$ is the dimensional regularization scale,
$\square_i\left( l\right)=l^2-m^{2}_{i}$ and
\begin{equation}
I_{G}(p^2)=\int_{0}^{1}dx\log\left[ 1-\frac{p^{2}}{m_{K}^{2}}%
x(1-x)-i\varepsilon\right] =-2+\sigma (p^{2})\log\frac{\sigma(p^{2})+1}{\sigma(p^{2})-1}.
\end{equation}
Here $\sigma(p^{2})=\sqrt{1-\frac{4m_{K}^{2}}{p^{2}}}$ and the integrals in
Eqs.(\ref{Idecay}, \ref{Jdecay}) are given by
\begin{eqnarray}
I_{P}& =&\int_{0}^{1}dx\int_{0}^{x}dy\frac{y(1-x)}{f_1(x,y)} , \qquad
J_{P} =\frac{1}{2}\int_{0}^{1}dx\int_{0}^{x}dy\frac{y(1-2y)}{f_1(x,y)}, \nonumber\\
J_{V} & =&\int_{0}^{1}dx\int_{0}^{x}dy\frac{y(1-x)}{f_2(x,y)},\qquad
I_{2} =\int_{0}^{1}dx\int_{0}^{x}dy\log[f_2(x,y)],  \label{I2} \nonumber\\
f_1(x,y)& =&1-\frac{Q^{2}}{m_{K}^{2}}%
x(1-x)+\frac{2Q\cdot k}{m_{K}^{2}}(1-x)y-\frac{k^{2}}{m_{K}^{2}}%
y(1-y)-i\varepsilon,  \nonumber\\
f_2(x,y)& =&f_1(x,y)-\frac{%
\left( m_{K^{\ast}}^{2}-m_{K}^{2}\right) }{m_{K}^{2}}(y-x).
\end{eqnarray}

The divergences in the loop integrals are contained in $a(\mu)$ in
Eqs. (\ref{Idecay},\ref{GKreg}) and we discuss its physical value in detail below.

From Eqs. (\ref{finalamp}, \ref{Jdecay}) we can see that the
dynamics is dominated by two main effects in the
$\gamma^{*}\phi\pi^{0}\eta$ vertex function which occur at two
different energy scales: the leading order terms for the
$K\overline{K}-\pi^0\eta$ on-shell scattering amplitude at the
di-meson invariant mass scale, $m_{\pi\eta}$, and the
electromagnetic meson form factors at the energy of the reaction, $
\sqrt{s}$. The calculation of the $\gamma^{*}\phi\pi^{0}\eta$ vertex
function involved in this amplitude is strictly accurate for low
di-meson invariant mass and low virtuality of the photon. We improve
this results in two respects: i) we replace the leading order result
for the $K\overline{K}-\pi^0\eta$ on-shell scattering amplitude by
the unitarized amplitude containing the scalar poles and ii) we
replace the leading order terms in the kaon form factor by the
unitarized one and, following \cite{GomezAvila:2007ru}, the leading
order terms of the $K^{*}K$ transition form factor are replaced by
the complete form factor at the energy of the reaction as extracted
from data.

The isovector $s$-wave $K\overline{K}-\pi^0\eta$ unitarized scattering amplitude denoted $t_{KK\pi\eta}^{0}$ gives
a successful description of the corresponding cross section up to $m_{\pi\eta} \approx 1 $GeV and
is based on the imposition of unitary constraints in coupled channels of
$\chi$PT. Following \cite{OO,OOPel}, unitarization reduces to the solution of the Bethe-Salpeter equation
\begin{equation}
\Te = \Te_{(2)} + \Te_{(2)} \cdot \mathbb{G} \cdot \Te, \label{LS}
\end{equation}
where $\Te$ is the matrix of unitarized amplitudes of the desired
isospin channel
\begin{equation}
\Te =\left(\begin{array}{cc}
t_{KKKK}^{0} & t_{KK\pi\eta}^{0}\\
t_{KK\pi\eta}^{0} & t_{\pi\eta\pi\eta}^{0}
     \end{array}\right).
\end{equation}
The elements of the matrix $\Te_{(2)}$ are the corresponding on-shell amplitudes
calculated in $\chi$PT at $\Opd$. A straightforward calculation yields
\begin{eqnarray}
\Te_{(2)} &=&\left(\begin{array}{cc}
V_{KKKK} & V_{KK\pi\eta}\\
V_{KK\pi\eta} & V_{\pi\eta\pi\eta}
     \end{array}\right) \nonumber \\
     &=&\left(\begin{array}{cc}
-\frac{m_{\pi\eta}^2}{4f^2} & \frac{9 m_{\pi\eta}^2-8 m_K^2-3 m_{\eta }^2-m_{\pi }^2}{6 \sqrt{6} f^2}\\
\frac{9 m_{\pi\eta}^2-8 m_K^2-3 m_{\eta }^2-m_{\pi }^2}{6 \sqrt{6} f^2}& -\frac{m_{\pi}^2}{3f^2}
     \end{array}\right).
\end{eqnarray}
The diagonal matrix $\mathbb{G}$ contains the loop integrals
\begin{equation} \label{Gnn}
\mathbb{G}=\left(\begin{array}{cc}
G_{KK}(m_{\pi\eta}^2) & 0\\
0 & G_{\pi\eta}(m_{\pi\eta}^2)
     \end{array}\right)
\end{equation}
with $G_{KK}(m_{\pi\eta}^2)$ given in Eq. (\ref{GKreg}) and
\begin{eqnarray}
G_{\pi\eta}(m_{\pi\eta}^{2})&=&\mu^{2\varepsilon}\int\frac{d^{d}l}{(2\pi)^{d}}\frac {i%
}{\square_{\pi}\left( l+Q\right) \square_{\eta}\left( l-k\right) } \nonumber \\
&=& \frac{1}{\left(
4\pi\right) ^{2}}\left[ a(\mu)+\log\frac{m_{\eta}^{2}}{\mu^{2}}+I_{\eta}(m_{\pi
\eta}^{2})\right]  \label{Geta}
\end{eqnarray}
with
\begin{eqnarray}
I_{\eta}&=&-2+\frac{m_{\pi\eta}^{2}-m_{\eta}^{2}+m_{\pi}^{2}}{2m_{\pi\eta}^{2}}\log\left(\frac{m_{\pi}^{2}}{m_{\eta}^{2}} \right) \nonumber \\
&&+\frac{\nu}{2m_{\pi\eta}^{2}}\left\{\log\left[\frac{\left(m_{\pi\eta}^{2}+\nu\right)^2-\left(m_{\eta}^{2}-m_{\pi}^{2}\right)^2}{\left(m_{\pi\eta}^{2}-\nu\right)^2-\left(m_{\eta}^{2}-m_{\pi}^{2}\right)^2}\right]-2\pi i\right\}.
\end{eqnarray}
and
\begin{equation}
 \nu=\sqrt{\left[m_{\pi\eta}^{2}-(m_\eta-m_\pi)^2\right]
 \left[m_{\pi\eta}^{2}-(m_\eta+m_\pi)^2\right]}.
\end{equation}

The substraction constant has been fixed in analogy with Ref. \cite{OOPel} to $%
a(\mu_{0})=0.87$ for $\mu_{0}=1.2$ GeV matching Eq. (\ref{Geta}) to the cutoff regularized
integral with a cutoff parameter $\Lambda=1$ GeV. It is related at different scales
as $a(\mu)=a(\mu _{0})+\log\frac{\mu^{2}}{\mu_{0}^{2}}$, and therefore
the loop function is scale independent.

\begin{figure}[ptb]
\begin{center}
\includegraphics[width=0.6\textwidth]{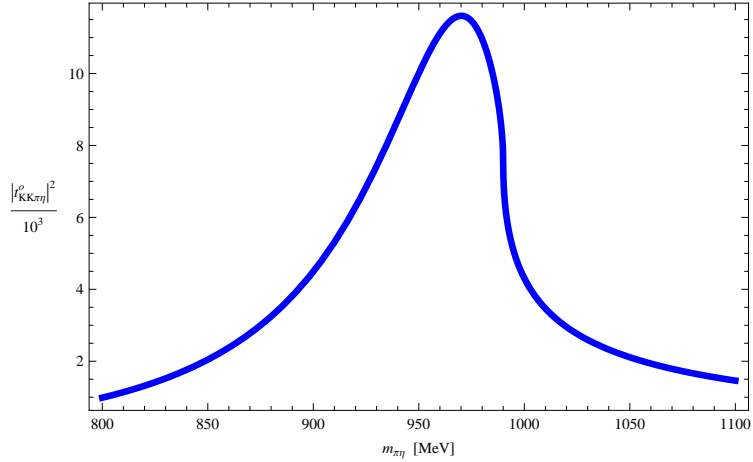}
\end{center}
\caption{Square modulus of the unitarized amplitude $t_{KK\pi\eta}^{0}$. }
\label{a0}
\end{figure}

The unitarized amplitudes are obtained solving the {\it algebraic} equation (\ref{LS}).
In particular, it was shown in \cite{OO} that $t_{KK\pi\eta}^{0}$ has a pole
corresponding to the $a_0$(980) resonance which manifests as a peak in the
squared amplitude as shown in  Fig. (\ref{a0}).

Concerning the meson electromagnetic form factors, the high virtuality of the
intermediate photon involved in our process requires to work out the
complete $\gamma K\overline{K}$ vertex functions. The calculation of the isovector kaon form factor ($F_{K}^{(1)}$) has been done in the context of $U\chi PT$ and the result can be
found in Eq.(31) of Ref. \cite{OOP} (see the discussion below this equation).  On the other hand, the extraction of the isovector
$K^{\ast}K$ transition form factor ($F_{K^{\ast }K}^{(1)}$) follows from \cite{GomezAvila:2007ru}, where an appropriate characterization of this form factor at the energy of the reaction is done.  As this form factor is physically dominated by resonances, in \cite{GomezAvila:2007ru} it is described with the lowest order terms obtained in R$\chi$PT complemented with the exchange of a $\rho^{\prime}$
\begin{align}
F_{K^{\ast}K}^{(1)}(k^{2}) &  =\frac{F_{V}\ G}{3}\left(  \frac{3m_{\rho}}%
{k^{2}-m_{\rho}^{2}}\right)  +b_{1}\left(  \frac{3m_{\rho^{\prime}}^{2}}%
{k^{2}-m_{\rho^{\prime}}^{2}+i \sqrt{s}\, \Gamma_{\rho^{\prime}}%
(s)}\right)  ,\label{KsKTFFI1}%
\end{align}
with the energy dependent width used in \cite{BBKsFF}
\begin{equation}
\Gamma_{\rho^{\prime}}(s)  =\Gamma_{\rho^{\prime}}\left[  \frac
{\mathcal{P}_{4\pi}(s)}{\mathcal{P}_{4\pi}(m_{\rho^{\prime}}^{2})}B_{4\pi
}^{\rho^{\prime}}+(1-B_{4\pi}^{\rho^{\prime}})\right]  ,
\end{equation}
where%
\begin{equation}
\mathcal{P}_{4\pi}(s)=\frac{\left(  s-16m_{\pi}^{2}\right)
^{3/2}}{s}.
\end{equation}
The values for the constants appearing here are extracted from the central
values of Table XV in \cite{BBKsFF} as $B_{4\pi}^{\rho^{\prime}}=0.65$,
$m_{\rho^{\prime}}=1504$ MeV and $\Gamma_{\rho^{\prime}}=438$ MeV
(see also \cite{DM2KsFF}). The
parameter $b_{1}$ is then fitted to the  experimental data for the
isovector cross sections reported for $e^{+}e^{-}\rightarrow K^{\ast}K$ at
$\sqrt{s}=$ $1400-3000\ $MeV in table VII of \cite{BBKsFF}. The
fit to the isovector cross section within $1\sigma$ yields
$b_{1}=-(0.255^{+0.030}_{-0.040})\times 10^{-3}$ MeV$^{-1}$ .

We use these results to calculate the
$\pi^0\eta$ spectrum, obtaining

\begin{equation}
\frac{d\sigma}{dm_{\pi\eta}}=\frac{\alpha^{2}}{24\pi^{5}m_{K}^{4}}\frac{|%
\mathbf{Q}||\widetilde{\mathbf{p}}|}{s^{\frac{3}{2}}}\
\frac{|t_{KK\pi\eta }^{0}|^2}{2}h(s,m_{\pi\eta}) \label{spectrum}
\end{equation}
where%
\begin{equation}
h(s,m_{\pi\eta})=|I|^{2}\,\left( M_{\phi}^{2}+2{\omega}^{2}\,\,\right) -\,6%
 Re(IJ^{\ast}\,)\sqrt{s}\,\omega\,+|J|^{2}\frac{s}{\,M_{\phi}^{2}}%
\,\left( 2M_{\phi}^{2}+{\omega}^{2}\right) .
\end{equation}
Here $(\omega,\mathbf{Q})$ stands for the momentum of the $\phi$ in the center of
momentum system of the reaction and $\widetilde{\mathbf{p}}$ denotes the
momentum of the final pion in the $\pi^0\eta$ center of momentum system%
\begin{equation}
\omega=\frac{s+M_{\phi}^{2}-m_{\pi\eta}^{2}}{2\sqrt{s}},\quad |\mathbf{Q}|=\frac{\lambda^{\frac{1}{2}}(s,M_{\phi}^{2},m_{\pi\eta}^{2})}{2%
\sqrt{s}},\quad
|\widetilde{\mathbf{p}}|=\frac{\lambda^{\frac{1}{2}%
}(m_{\pi\eta}^{2},m_{\pi}^{2},m_{\eta}^{2})}{2m_{\pi\eta}}.
\end{equation}
with
\begin{equation}
\lambda(m_{1}^{2},m_{2}^{2},m_{3}^{2})=(m_{1}^{2}-(m_{2}-m_{3})^{2})(m_{1}%
^{2}-(m_{2}+m_{3})^{2}).
\label{lambda}
\end{equation}
\begin{center}
\begin{figure}[ptb]
\begin{center}
\includegraphics[
width=0.6 \textwidth]{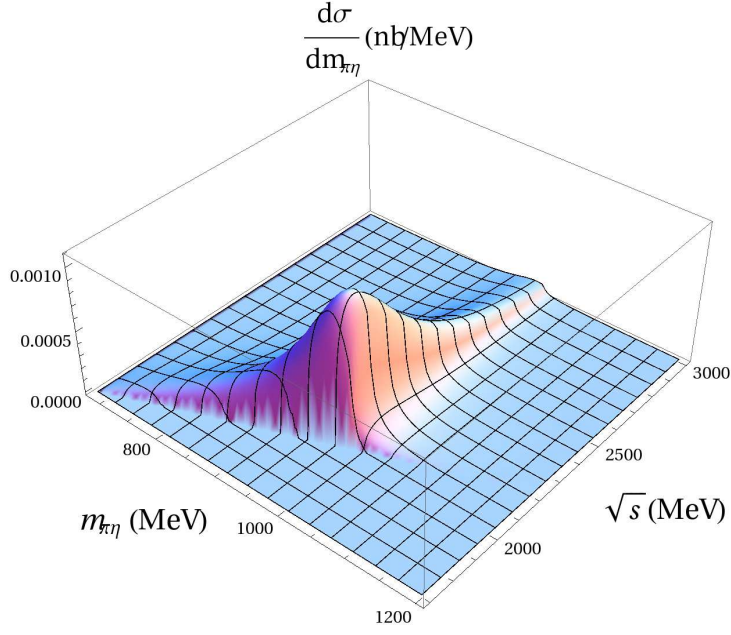}
\end{center}
\caption{Differential cross section as a function of the $\pi^0\eta$
invariant mass and of the center of mass energy.}
\label{dsig}
\end{figure}
\end{center}

\section{Numerical results}
\begin{center}
\begin{figure}[ptb]
\begin{center}
\includegraphics[
width=0.5 \textwidth]{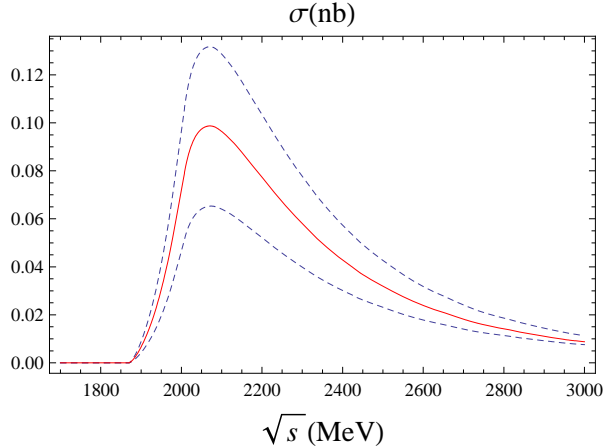}
\end{center}
\caption{Cross section for
$e^{+}e^{-}\rightarrow\phi\left[\pi^0\eta\right] _{I=1,J=0}$ as a function
of $\sqrt{s}$. The dashed lines correspond to the $1\sigma$ region in the
extraction of the isovector $K^{\ast}K$ transition form factor in
Ref. \cite{GomezAvila:2007ru}, i.e.  to the fit of $b_1$ in Eq. (\ref{KsKTFFI1}).}
\label{sig}
\end{figure}
\end{center}

We evaluate numerically the integrals and the differential cross
section. Using the physical masses and coupling constants
$m_{K}=495\, \text{MeV}$, $m_{\phi}=1019.4\, \text{MeV}$,
$\alpha=1/137$, $G_{V}=53\, \text{MeV}$, $F_{V}=154\, \text{MeV}$,
$f_{\pi}=93 \, \text{MeV}$ , and $G=0.016\, \text{MeV}^{-1}$ in Eq.
(\ref{spectrum}) we obtain the spectrum shown in Fig. (\ref{dsig})
where the presence of the $a_{0}(980)$ is well visible. This is a
consequence of the fact that the $a_{0}(980)$ poles are well
reproduced in the unitarization of meson-meson $s$-wave isovector
amplitudes present in our calculation. Next we integrate $m_{\pi
\eta }$ in the $a_0(980)$ region, $m_{\pi\eta }=850-1100
\,\text{MeV}$. The obtained cross section is shown in Fig
(\ref{sig}).

The mechanisms studied here were shown to be responsible for the
production of most of the events in the case of the $\phi\pi\pi$
final state  except for the resonant events due to the $X(2175)$
\cite{Napsuciale:2007wp}. On the other hand, in the case of the
$\phi K^{+}K^{-}$ final state studied in \cite{GomezAvila:2007ru},
within the limitations due to the extraction of the $K^{\ast }K$
isovector transition form factor from data, a good description of
experimental points is obtained but there seems to be room for
additional contributions around $2200\, \text{MeV}$. In this case,
the $\phi K^{+}K^{-}$ system can be in both isoscalar and isovector
states hence the natural candidate for additional contributions is
an isovector companion of the $X(2175)$. Interestingly, in the
three-body description of the $X(2175)$ proposed in
\cite{MartinezTorres:2008gy}, no peak is generated in the isovector
channel. Our calculation of the pure isovector channel
$e^{+}e^{-}\rightarrow\phi a_0(980)$ under the mechanisms studied in
\cite{Napsuciale:2007wp} yields a cross section of the same size as
that of $e^+e^-\to \phi f_0$ and thus within the reach of present
$e^{+}e^{-}$ machines. If there is no such a thing as an isovector
companion of the $X(2175)$ as expected from  \cite{MartinezTorres:2008gy},
then our result is a concrete theoretical prediction for the
$e^{+}e^{-}\rightarrow\phi a_0(980)$ cross section; otherwise, the
isovector companion  must contribute to this process yielding valuable
information on the nature of mesons at this
energy. Hence we encourage experimentalists to measure the
$e^{+}e^{-}\rightarrow\phi \pi^{0}\eta$ channel.\\

\noindent {\bf Acknowledgments}
This work was supported by CONACyT-M\'{e}xico under project
CONACyT-50471-F. M.N. acknowledges partial support from DINPO-UG.

\end{document}